\documentclass[12pt]{article}

\author{Yu.~M.~Zinoviev
       \thanks{E-mail address: Yurii.Zinoviev@ihep.ru} \\[0.5cm]
        {\it Institute for High Energy Physics} \\
        {\it of National Research Center "Kurchatov Institute"} \\
        {\it Protvino, Moscow Region, 142280, Russia}}

\title{On hypersymmetry in three dimensions}

\date{}

\begin{document}

\maketitle

\begin{abstract}
In this work we presented a number of explicit examples for the cubic
vertices describing an interaction of massless spin-$\frac{5}{2}$
field with massive boson and fermion including all
hypertransformations necessary for the vertices to be gauge invariant.
Here we restrict ourselves with the massive bosons with spins
$s=2,1,0$ and massive fermions with spins $s=\frac{3}{2},\frac{1}{2}$.
Our general analysis predicted that the vertex must exist for any
boson and fermion with the spin difference $\frac{3}{2}$ or 
 $\frac{1}{2}$. And indeed it appeared that the vertex exists for all
six possible pairs $(2,1,0) \otimes (\frac{3}{2},\frac{1}{2})$. As in
the case of massive supermultiplets, our construction is based on the
gauge invariant description for the massive fields with spins $s \ge
1$. Moreover, we have explicitly checked that all the vertices are
invariant also under the gauge symmetries of these massive fields.

\end{abstract}

\thispagestyle{empty}
\newpage
\setcounter{page}{1}

\section{Introduction}

After the discovery of supersymmetry and supergravity in 70-ies
naturally there was a number of attempts to consider even higher
symmetries one of the next candidate being hypersymmetry with the spin
$\frac{5}{2}$ as a gauge field. All such attempts failed and now,
looking back, we understand why: in four dimensions any theory
containing field with spin $s > 2$ must unavoidably include the whole
infinite set of them  (see e.g. \cite{BBS10} and references therein).

Happily, in three dimensions many things become much simpler and
in-particular there exist a number of closed self-consistent models
with a finite number of higher spin fields. Not surprisingly, a
hypergravity appeared in three dimensions \cite{AD84} as a natural
generalization of the three dimensional supergravity. In flat three
dimensional space an interaction of gravity with gravitino has a
remarkably simple form (especially if one uses a frame-like
multispinor formalism where gravity is described by one-forms
$H^{\alpha(2)}$ and $\Omega^{\alpha(2)}$ while gravitino --- by
one-form $\Phi^\alpha$):
$$
{\cal L}_{int} = g_0 \Phi_\alpha \Omega^\alpha{}_\beta \Phi^\beta,
\qquad \delta H^{\alpha(2)} \sim g_0 \Phi^\alpha \zeta^\alpha. 
$$
A technical reason for such simplicity is that the interaction vertex
contain only Lorentz connection $\Omega^{\alpha(2)}$ while non-trivial
supertransformations  acquires only physical field $H^{\alpha(2)}$.
As a result there are now any quadratic variations and hence there is
no need to introduce any other fields or consider something like
four-fermion interaction. As it was noted in \cite{AD84} this
mechanism nicely work for the spin $\frac{5}{2}$ field
$\Phi^{\alpha(3)}$ as well:
$$
{\cal L}_{int} = g_0 \Phi_{\alpha(2)\beta} \Omega^\beta{}_\gamma
\Phi^{\alpha(2)\gamma}, \qquad \delta H^{\alpha(2)} \sim g_0
\Phi^{\alpha\beta(2)} \zeta^\alpha{}_{\beta(2)}. 
$$
Moreover, this construction really works for the arbitrary
half-integer spin $\Phi^{\alpha(2n+1)}$:
$$
{\cal L}_{int} = g_0 \Phi_{\alpha(2n)\beta} \Omega^\beta{}_\gamma
\Phi^{\alpha(2n)\gamma}, \qquad \delta H^{\alpha(2)} \sim g_0
\Phi^{\alpha\beta(2n)} \zeta^\alpha{}_{\beta(2n)}. 
$$
However, if one tries to deform these models into $AdS_3$ background
one immediately find that a non-zero cosmological constant spoils the
invariance under the fermionic transformations. The technical reason
is again clear: the interaction vertices contain now both 
$\Omega^{\alpha(2)}$ and $H^{\alpha(2)}$ and both of them acquire
non-trivial transformations. Thus there appear quadratic variations
which cancel each other in the supergravity but not in its higher spin
generalizations. This in turn may signal that our models are
incomplete and there are some missing components.

Both gravity and supergravity in $AdS_3$ can be compactly rewritten as
the so-called Chern-Simons theories \cite{AT86,Wit88}. For the gravity
this based on the isomorphism:
$$
SO(2,2) \approx SL(2) \otimes SL(2) \approx Sp(2) \otimes Sp(2).
$$
As a result, the Lagrangian can be written as a difference of two
equivalent parts
$$
{\cal L} = {\cal L}(\Omega_+) - {\cal L}(\Omega_-), \qquad
\Omega_{\pm}^{\alpha(2)} = \Omega^{\alpha(2)} \pm \frac{\lambda}{2}
H^{\alpha(2)}, \qquad \lambda = \sqrt{- \Lambda}, 
$$
$$
{\cal L}(\Omega) = \frac{1}{2} \Omega_{\alpha(2)} D \Omega^{\alpha(2)}
+ \frac{g_0}{3} \Omega_\alpha{}^\beta \Omega_\beta{}^\gamma
\Omega_\gamma{}^\alpha 
$$
each one invariant under its own gauge transformations. As far as the
bosonic fields are concerned, this construction admits a lot of
possible generalizations to higher spins \cite{Vas91,CFPT10}. A simple
example is a class of models based on the algebras 
$SL(N) \otimes SL(N)$ describing a set of massless fields with spins 
$2,3,4,\dots,N$. Let us stress that it is crucial that there exists a
so-called canonical embedding $SL(2) \in SL(N)$  so that the gauge
fields corresponding to the generators of $SL(N)$ can be interpreted
as the higher spin ones. Later on it was understood that it is possible
to truncate such models to even spins $2,4,\dots,2n$ only with the
gauge algebra being $Sp(2n) \otimes Sp(2n)$ \cite{CLW12}. 

Let us return back to the fermions. It was already mentioned that the
supergravity admits deformation into $AdS_3$, so it is worth to look
at it more closely. As it was shown in \cite{AT86} the most general
case corresponds to the Chern-Simons theory based on the superalgebra
$OSp(M,2) \otimes OSp(N,2)$. Moreover if one does not insist on parity
conservation these two integers $M$ and $N$ may be different. Let us
consider a simple case of a so-called $(1,1)$ supergravity with the
superalgebra $OSp(1,2) \otimes OSp(1,2)$. AS in the bosonic case, the
Lagrangian can be written as
$$
{\cal L} = {\cal L}(\Omega_+, \Phi_+) - {\cal L}(\Omega_-, \Phi_-),
$$
where each part has the form:
$$
{\cal L}(\Omega, \Phi) = \frac{1}{2} \Omega_{\alpha(2)} D
\Omega^{\alpha(2)} + \frac{i}{2} \Phi_\alpha D \Phi^\alpha 
+ \frac{g_0}{3} \Omega_\alpha{}^\beta \Omega_\beta{}^\gamma
\Omega_\gamma{}^\alpha + \frac{ig_0}{2} \Phi_\alpha 
\Omega^\alpha{}_\beta \Phi^\beta 
$$
and is invariant under the following gauge transformations:
\begin{eqnarray*}
\delta \Omega^{\alpha(2)} &=& D \eta^{\alpha(2)} + g_0 
\Omega^\alpha{}_\beta \eta^{\alpha\beta} + \frac{ig_0}{2}
\Phi^\alpha \zeta^\alpha, \\
\delta \Phi^\alpha &=& D \zeta^\alpha + g_0 \Omega^\alpha{}_\beta
\zeta^\beta + g_0 \eta^{\alpha\beta} \Phi_\beta. 
\end{eqnarray*}
Surprisingly, as it was noted in \cite{Zin14a}, these simple formulas
really describe not one theory, but an infinite number of different
ones including hypergravity in $AdS_3$. All one has to do is to
replace spinor index $\alpha=1,2$ to some index $a=1,2,\dots,2n$ and
the anti-symmetric bi-spinor $\epsilon^{\alpha\beta}$ by a completely
anti-symmetric non-degenerate invariant tensor $\epsilon^{ab}$, 
$\epsilon_{ab} \epsilon^{bc} = - \delta_a{}^c$.

Thus the hypergravity in $AdS_3$ can be written as a Chern-Simons
theory with the superalgebra $OSp(1,4) \otimes OSp(1,4)$. Recall that
the superalgebra $OSp(1,4)$ is very well known as the superalgebra of
$AdS_4$ supergravity, but now it appears as the algebra of $AdS_3$
hypergravity. In this case it is also crucial that there exists a
canonical embedding $Sp(2) \in OSp(1,4)$ where the index $a=1,2,3,4$
(which in $d=4$ is just an index of Majorana spinor) becomes a
completely symmetric tri-spinor $\alpha(3)$. So for the gauge fields
we obtain:
\begin{eqnarray*}
\Omega^{(ab)} &=& \Omega^{(\alpha(3),\beta(3))} 
= \Omega^{\alpha(3)\beta(3)} + \epsilon^{\alpha\beta}
\epsilon^{\alpha\beta} \Omega^{\alpha\beta}, \\
\Phi^a &=& \Phi^{\alpha(3)}.
\end{eqnarray*}
Thus the hypergravity must contain bosonic fields with spins $s=2,4$
and fermionic one with spin $s=5/2$ (see \cite{Rah19} for the proof of
the uniqueness). As in the flat case, this construction admits a
generalization to the superalgebra $OSp(1,2n) \otimes OSp(1,2n)$
describing a set of even integer spins $2,4, \dots,2n$ and just one
half-integer spin $\frac{(2n+1)}{2}$ . 
 
A very important part of any study of the gravity and supergravity is
their interactions with matter, where by matter one usually means
massive fields with spins (0,1/2,1). For the higher spins in $d=3$
till now there exists only one non-trivial example ---
Prokushkin-Vasiliev theory \cite{PV98} (see also \cite{Ble89}) which
describes interactions of the infinite set of massless higher spins
with massive scalars and spinors. But in a broader sense one can
consider any massive (even higher spin) fields as "matter" and this
raises the question on the general massless to massive fields
interactions. Recently \cite{Zin21} we have considered the most
general cubic vertices for the two arbitrary (bosonic or fermionic)
fields $(M_1,s_1)$ and $(M_2,s_2)$ with the massless field $(0,s_3)$.
We found that for the vertex to exist masses and spins must satisfy:
$$
M_1 - M_2 = (s_1-s_2)\lambda, \qquad
s_i + s_{i+1} > s_{i+2}.
$$
In-particular, in the flat limit two masses must be equal. As for the
spins, they must satisfy a so-called triangular inequality. To
illustrate how this inequality works let us consider some simple
examples.
\begin{itemize}
\item Supergravity. For the massless $s=\frac{3}{2}$ the spins of the
massive boson and fermion must differ by $\frac{1}{2}$ so they form a
massive supermultiplet (see \cite{BSZ17a,Zin23}).
\item (Non-)gravity. For the massless spin-2 usually associated with
gravity there exist interactions where spins of two bosons or two
fermions differ by 1 \cite{Zin22}.
\item Hypergravity. For the massless $s=\frac{5}{2}$ the spins of the
massive boson and fermion may differ by $\frac{1}{2}$ or by 
$\frac{3}{2}$.
\end{itemize}

Out aim in this work is a direct construction of several examples of
the cubic vertices describing interactions for the massless 
spin-$\frac{5}{2}$ field with massive bosons and fermions. We restrict
ourselves with the massive bosons with spins $s=2,1,0$ and the massive
fermions with spins $s=\frac{3}{2},\frac{1}{2}$. It is easy to see
that there exist six pairs satisfying the triangular inequality and it
appears that in all six cases the corresponding vertices do exist. To
make these vertices to be gauge invariant we introduce all necessary
hypertransformations for all fields. This in turn guarantee that the
corresponding hypercurrents are conserved on-shell. One of the
technical problems here is that lower spin components are zero-forms,
while all higher spin ones are one-forms. So one faces the situations
when some zero-form must transform into one-form. In all such
situations we switch to the metric-like multispinor formalism (see
below) similar to the one used in \cite{KT16,KP18,KhZ22,Zin23}. Let us
stress that all cubic vertices constructed in our work are written in
the pure frame-like formalism, while the metric-like one is used only
to write the hypertransformations for the zero-forms as well as at
some intermediate steps of calculations.

The paper is organized as follows. In section 2 we provide all
necessary kinematical information for massive fields with spins
$s=2,\frac{3}{2},1,\frac{1}{2},0$. Then in section 3 we present the
cubic vertices and hypertransformations for all six cases considered.

\noindent
{\bf Notation and conventions} We work in the frame-like multispinor
formalism where all fields are one-forms or zero-forms having a set of
completely symmetric local spinor indices $\alpha=1,2$. For coordinate
free description of the flat three dimensional space we use background
frame $e^{\alpha(2)}$ and background Lorentz covariant derivative $D$
such that
\begin{equation}
D \wedge D = 0, \qquad D \wedge e^{\alpha(2)} = 0. 
\end{equation}
Besides $e^{\alpha(2)}$, a complete basis of forms contains two-form
$E^{\alpha(2)}$ and three-form $E$ defined as follows
\begin{equation}
e^{\alpha(2)} \wedge e^{\beta(2)} = \epsilon^{\alpha\beta}
E^{\alpha\beta}, \qquad E^{\alpha(2)} \wedge e^{\beta(2)} =
\epsilon^{\alpha\beta} \epsilon^{\alpha\beta} E.
\end{equation}
As it was already mentioned, each time as it is convenient we switch
to the metric-like multispinor formalism using the fact that any 
one-form can be transformed into a set of zero-forms (see
\cite{KhZ22} for details), for example
\begin{eqnarray*}
H^{\alpha(2)} &=& e_{\beta(2)} h^{\alpha(2)\beta(2)} + 
e^\alpha{}_\beta h^{\alpha\beta} + e^{\alpha(2)} h, \\
\Phi^\alpha &=& e_{\beta(2)} \phi^{\alpha\beta(2)} + e^\alpha{}_\beta
\phi^\beta. 
\end{eqnarray*}

\section{Kinematics}

In this section we provide all necessary kinematical information for
all massive fields entering the cubic vertices we consider. Recall
that for the construction of massive supermultiplets
\cite{BSZ17a,Zin23} it was crucial to use a gauge invariant
description for the massive fields with $s \ge 1$. The same holds true
for the hypersymmetry as well.

\subsection{Spin 0}

Here we use a first order formalism with the physical field $\varphi$
and the auxiliary one $\pi^{\alpha(2)}$. The free Lagrangian has the
form:
\begin{equation}
{\cal L}_0 = - \frac{1}{2} \pi_{\alpha(2)} \pi^{\alpha(2)} -
\pi_{\alpha(2)} D^{\alpha(2)} \varphi - \frac{M^2}{4} \varphi^2,
\end{equation}
while the Lagrangian equations look like:
\begin{equation}
0 = - \pi^{\alpha(2)} - D^{\alpha(2)} \varphi, \qquad
0 = (D\pi) - \frac{M^2}{2} \varphi.
\end{equation}
To simplify calculations of the cubic vertices in what follows we work
up to the terms proportional to the auxiliary field
$\pi^{\alpha(2)}$-equation. This means, in-particular:
\begin{equation}
D^\alpha{}_\beta \pi^{\alpha\beta} \approx 0, \qquad
D^{\alpha(2)} \pi^{\beta(2)} \approx \frac{1}{6} 
\pi^{\alpha(2)\beta(2)} + \frac{1}{12} \epsilon^{\alpha\beta}
\epsilon^{\alpha\beta} (D\pi), 
\end{equation}
where
$$
\pi^{\alpha(4)} = D^{\alpha(2)} \pi^{\alpha(2)}.
$$

\subsection{Spin $\frac{1}{2}$}

In the multispinor formalism the free Lagrangian has the form
\begin{equation}
{\cal L}_0 = \frac{1}{2} \rho_\alpha D^\alpha{}_\beta \rho^\beta
- \frac{M_0}{4} \rho_\alpha \rho^\alpha,
\end{equation}
while Lagrangian equation look like:
\begin{equation}
0 = D^\alpha{}_\beta \rho^\beta - \frac{M_0}{2} \rho^\alpha.
\end{equation}
In what follows we use
\begin{equation}
D^{\alpha(2)} \rho^\beta = \frac{1}{3} \rho^{\alpha(2)\beta}
+ \frac{1}{3} \epsilon^{\alpha\beta} D^\alpha{}_\gamma \rho^\gamma,
\qquad \rho^{\alpha(3)} = D^{\alpha(2)} \rho^\alpha. 
\end{equation}

\subsection{Spin 1}

In this case we use a first order gauge invariant formalism with the
physical fields $A^{\alpha(20}$, $\varphi$ and the auxiliary ones
$B^{\alpha(2)}$, $\pi^{\alpha(2)}$. The free Lagrangian has the form
\begin{equation}
{\cal L}_0 = \frac{1}{2} B_{\alpha(2)} B^{\alpha(2)} + 4
B^{\alpha\beta} D_\alpha{}^\gamma A_{\beta\gamma} - \frac{1}{2}
\pi_{\alpha(2)} \pi^{\alpha(2)} - \pi_{\alpha(2)} D^{\alpha(2)}
\varphi + 2M \pi^{\alpha(2)} A_{\alpha(2)}
\end{equation}
and is invariant under the following gauge transformations
\begin{equation}
\delta A^{\alpha(2)} = D^{\alpha(2)} \xi, \qquad
\delta \varphi = 2M \xi.
\end{equation}
A complete set of the Lagrangian equations looks like:
\begin{eqnarray}
0 &=& B_{\alpha(2)} + 2 D_\alpha{}^\beta A_{\alpha\beta}, \qquad
0 = 2 D_\alpha{}^\beta B_{\alpha\beta} + 2M \pi_{\alpha(2)}, \\
0 &=& - \pi_{\alpha(2)} - D_{\alpha(2)} \varphi + 2M A_{\alpha(2)},
\qquad 0 = (D\pi).
\end{eqnarray}
Again, in what follows we work up to the terms proportional to the
auxiliary fields  $B^{\alpha(2)}$ and $\pi^{\alpha(2)}$ equations, so
that
\begin{equation}
(DB) \approx 0, \qquad 
D^\alpha{}_\beta \pi^{\alpha\beta} \approx 2M D^\alpha{}_\beta
A^{\alpha\beta} \approx M B^{\alpha(2)}.
\end{equation}
In-particular, we use
\begin{eqnarray}
D^{\alpha(2)} B^{\beta(2)} &\approx& \frac{1}{6} B^{\alpha(2)\beta(2)}
+ \frac{1}{8} \epsilon^{\alpha\beta} (D^\alpha{}_\gamma
B^{\beta\gamma} + D^\beta{}_\gamma B^{\alpha\gamma}), \nonumber \\
D^{\alpha(2)} \pi^{\beta(2)} &\approx& \frac{1}{6} 
\pi^{\alpha(2)\beta(2)} + \frac{1}{12} \epsilon^{\alpha\beta}
\epsilon^{\alpha\beta} (D\pi) + \frac{M}{4} \epsilon^{\alpha\beta}
B^{\alpha\beta}, 
\end{eqnarray}
where
$$
B^{\alpha(4)} = D^{\alpha(2)} B^{\alpha(2)}.
$$

\subsection{Spin $\frac{3}{2}$}

The gauge invariant description of the massive spin-$\frac{3}{2}$
requires the one-form $\Phi^\alpha$ and zero-form $\rho^\alpha$. The
free Lagrangian has the form:
\begin{equation}
{\cal L}_0 = - \frac{1}{2} \Phi_\alpha D \Phi^\alpha + \frac{1}{2}
\rho_\alpha E^\alpha{}_\beta D \rho^\beta + 2M_0 \Phi_\alpha
E^\alpha{}_\beta \rho^\beta - \frac{M_0}{2} \Phi_\alpha
e^\alpha{}_\beta \Phi^\beta - \frac{3M_0}{2} E \rho_\alpha
\rho^\alpha.
\end{equation}
This Lagrangian is invariant under the following local
transformations
\begin{equation}
\delta \Phi^\alpha = D \zeta^\alpha + M_0 e^\alpha{}_\beta
\zeta^\beta, \qquad \delta \rho^\alpha = 2M_0 \zeta^\alpha. 
\end{equation}
Here the Lagrangian equations look like:
\begin{eqnarray}
0 &=& - D \Phi^\alpha - M_0 e^\alpha{}_\beta \Phi^\beta + 2M_0 
E^\alpha{}_\beta \rho^\beta, \nonumber \\
0 &=& E^\alpha{}_\beta D \rho^\beta - 2M_0 E^\alpha{}_\beta \Phi^\beta
- 3M_0 E \rho^\alpha \\
 &=& E [ 2 D^\alpha{}_\beta \rho^\beta + 6M_0 \phi^\alpha - 3M_0
\rho^\alpha], \nonumber
\end{eqnarray}
where the zero-form $\phi^\alpha$ comes from the decomposition
$$
\Phi^\alpha = e_{\beta(2)} \phi^{\alpha\beta(2)} + e^\alpha{}_\beta
\phi^\beta. 
$$

\subsection{Spin 2}

In the frame-like gauge invariant formalism massive spin-2 is
described by the free Lagrangian
\begin{eqnarray}
{\cal L}_0 &=& \Omega_{\alpha\beta} e^\beta{}_\gamma
\Omega^{\alpha\gamma} + \Omega_{\alpha(2)} D H^{\alpha(2)} 
 + E B_{\alpha(2)} B^{\alpha(2)} - (eB) D A - E \pi_{\alpha(2)}
\pi^{\alpha(2)} + (E\pi) D \varphi \nonumber \\
 && - 2M (e\Omega) A - M H_{\alpha\beta} E^\beta{}_\gamma
B^{\alpha\gamma} + 4M (E\pi) A \nonumber \\
 && + \frac{M^2}{4} H_{\alpha\beta} e^\beta{}_\gamma H^{\alpha\gamma}
+ M^2 (EH) \varphi + \frac{3}{2}M^2 E \varphi^2.
\end{eqnarray}
This Lagrangian is invariant under the following local
transformations
\begin{eqnarray}
\delta \Omega^{\alpha(2)} &=& D \eta^{\alpha(2)} + \frac{M^2}{4} 
e^\alpha{}_\beta \xi^{\alpha\beta}, \nonumber \\
\delta H^{\alpha(2)} &=& D \xi^{\alpha(2)} + e^\alpha{}_\beta
\eta^{\alpha\beta} + 2M e^{\alpha(2)} \xi, \nonumber \\
\delta B^{\alpha(2)} &=& 2M \eta^{\alpha(2)}, \qquad
\delta A = D \xi + \frac{M}{4} (e\xi), \\
\delta \pi^{\alpha(2)} &=& M^2 \xi^{\alpha(2)}, \qquad
\delta \varphi = - 4M \xi. \nonumber
\end{eqnarray}
The equations for the one-forms look like:
\begin{eqnarray}
0 &=& D \Omega^{\alpha(2)} - \frac{M}{2} E^\alpha{}_\beta
B^{\alpha\beta} + \frac{M^2}{4} e^\alpha{}_\beta H^{\alpha\beta}
+ M^2 E^{\alpha(2)} \varphi, \nonumber \\
0 &=& D H^{\alpha(2)} + e^\alpha{}_\beta \Omega^{\alpha\beta} 
+ 2M e^{\alpha(2)} A, 
\end{eqnarray}
while for for the zero-forms we obtain (up to three-form $E$ as a
common multiplier)
\begin{eqnarray}
0 &=& - 4 [ D^\alpha{}_\beta B^{\alpha\beta} + 8M \omega^{\alpha(2)}
- 2M \pi^{\alpha(2)} ], \nonumber \\
0 &=& 2 [ B^{\alpha(2)} - 2 D^\alpha{}_\beta A^{\alpha\beta}
- 2M h^{\alpha(2)}, \nonumber \\
0 &=& - [ 2 (D\pi) -6M^2 h - 3M^2 \varphi ], \\
0 &=& 2 [ - \pi^{\alpha(2)} + D^{\alpha(2)} \varphi + 4M
A^{\alpha(2)}]. \nonumber
\end{eqnarray}
Here and in what follows we use the decomposition
\begin{eqnarray*}
\Omega^{\alpha(2)} &=& e_{\beta(2)} \omega^{\alpha(2)\beta(2)}
+ e^\alpha{}_\beta \omega^{\alpha\beta} + e^{\alpha(2)} \omega, \\
H^{\alpha(2)} &=& e_{\beta(2)} h^{\alpha(2)\beta(2)} 
+ e^\alpha{}_\beta h^{\alpha\beta} + e^{\alpha(2)} h.
\end{eqnarray*}
In this case on the auxiliary fields $B^{\alpha(2)}$ and 
$\pi^{\alpha(2)}$ equations we have
\begin{equation}
(DB) \approx 2M (Dh), \qquad
D^\alpha{}_\beta \pi^{\alpha\beta} \approx 2M B^{\alpha(2)} - 4M^2
h^{\alpha(2)}.
\end{equation}

\section{Cubic vertices}

In this section we present a direct construction for six cubic
vertices describing an interaction of the massless spin-5/2 field with
the massive $s=2,1,0$ bosons and $s=\frac{3}{2},\frac{1}{2}$ fermions.
Our analysis in \cite{Zin21} based on the unfolded formulation has
shown that such vertices must contain terms with up to two
derivatives. Thus from the very beginning we restrict the number of
derivatives and in-particular forbid any field redefinitions raising
this number. 

\subsection{Vertex $(\frac{5}{2}-\frac{1}{2}-0)$}

The most general (up to possible field redefinitions) ansatz for the
terms with two derivatives (here and in what follows we take into
account that the auxiliary field $\pi^{\alpha(2)}$ is equivalent to
the first derivative of the physical field $\varphi$):
\begin{equation}
{\cal L}_{12} = \Psi_{\alpha(3)} [ g_1 e^{\alpha(2)} \pi^{\alpha\beta}
D \rho_\beta + g_2 e^{\alpha\beta} \pi^{\alpha(2)} D \rho_\beta
+ g_3 e^{\alpha\beta} \pi^\alpha{}_\beta D \rho^\alpha]. 
\end{equation}
However, there is an identity
$$
2 e^{\alpha(2)} \pi^{\alpha\beta} D \rho_\beta - 
2 e^{\alpha\beta} \pi^{\alpha(2)} D \rho_\beta
= - e^{\alpha\beta} \pi^\alpha{}_\beta D \rho^\alpha 
$$
and so we use only terms with the coefficients $g_{1,2}$. Variations
under the hypertransformations $\delta \Psi_{\alpha(3)} = D
\zeta_{\alpha(3)}$ produce
\begin{eqnarray}
\delta_0 {\cal L}_{12} &=& \zeta_{\alpha(3)} [ g_1 e^{\alpha(2)} D
\pi^{\alpha\beta} + g_2 e^{\alpha\beta} D \pi^{\alpha(2)} ] 
D \rho_\beta  \nonumber \\
 &=& - \zeta_{\alpha(3)} [ \frac{2g_1+g_2}{3} E^\alpha{}_\gamma
\pi^{\alpha(2)\beta\gamma} - 2\frac{g_1-g_2}{3} E^{\alpha(2)}
\epsilon^{\alpha\beta} (D\pi) + g_2 E^\beta{}_\gamma 
\pi^{\alpha(3)\gamma} ] D \rho_\beta. 
\end{eqnarray}
We put
$$
g_2 = - 2g_1
$$
and obtain (here and in what follows we omit the three form $E$ when
it appears as a common multiplier):
\begin{equation}
\delta_0 {\cal L}_{12} = \zeta_{\alpha(3)} [ 4g_1 \pi^{\alpha(3)\beta}
D_\beta{}^\gamma \rho_\gamma + 4g_1 (D\pi) \rho^{\alpha(3)}].
\end{equation}
Now we introduce hypertransformations for the scalar and spinor:
\begin{equation}
\delta_1 \rho^\alpha = - 4g_1 \pi^{\alpha\beta(3)} \zeta_{\beta(3)},
\qquad \delta_1 \varphi = 4g_1 \rho^{\alpha(3)} \zeta_{\alpha(3)} 
\end{equation}
and get\footnote{Here and in what follows ${\cal L}_0$ is the sum of
the two free Lagrangians corresponding to massive fields considered.}
\begin{equation}
\delta_0 {\cal L}_{12} + \delta_1 {\cal L}_0 = \zeta_{\alpha(3)} [ -
2M_0g_1 \pi^{\alpha(3)\beta} \rho_\beta + 2M^2g_1 \varphi
\rho^{\alpha(3)}].
\end{equation}
Then we add one-derivative terms:
\begin{equation}
{\cal L}_{11} = \Psi_{\alpha(3)} [ f_1 E^{\alpha(2)} \pi^{\alpha\beta}
+ f_2 E^{\alpha\beta} \pi^{\alpha(2)} ] \rho_\beta. 
\end{equation}
Their variations produce:
\begin{eqnarray}
\delta_0 {\cal L}_{11} &=& - \zeta_{\alpha(3)} [ (f_1 E^{\alpha(2)} D
\pi^{\alpha\beta} + f_2 E^{\alpha\beta} D \pi^{\alpha(2)}) \rho_\beta
+ (f_1 E^{\alpha(2)} \pi^{\alpha\beta} + f_2 E^{\alpha\beta}
\pi^{\alpha(2)}) D \rho_\beta] \nonumber \\
 &=& - 2 \zeta_{\alpha(3)} [ (f_1 D^{\alpha(2)} \pi^{\alpha\beta}
+ f_2 D^{\alpha\beta} \pi^{\alpha(2)}) \rho_\beta + f_1
\pi^{\alpha\beta} D^{\alpha(2)} \rho_\beta - f_2 \pi^{\alpha(2)}
D^\alpha{}_\beta \rho^\beta ]. 
\end{eqnarray}
Setting $f_1=0$ we obtain
\begin{equation}
\delta_0 {\cal L}_{11} = \zeta_{\alpha(3)} [ - f_2 
\pi^{\alpha(3)\beta} \rho_\beta + 2f_2 \pi^{\alpha(2)}
D^\alpha{}_\beta \rho^\beta].
\end{equation}
Then we introduce additional corrections for spinor
\begin{equation}
\delta_2 \rho^\alpha = 6f_2 \pi_{\beta(2)} \zeta^{\alpha\beta(2)}
\end{equation}
so that 
\begin{equation}
\delta_0 {\cal L}_{11} + \delta_2 {\cal L}_0 = \zeta_{\alpha(3)} [-
f_2 \pi^{\alpha(3)\beta} \rho_\beta + M_0f_2 \pi^{\alpha(2)}
\rho^\alpha ].
\end{equation}
At last, adding the only possible term without derivatives
\begin{equation}
{\cal L}_{10} = h_1 \Psi_{\alpha(3)} E^{\alpha(2)} \varphi \rho^\alpha
\end{equation}
we obtain its variations:
\begin{equation}
\delta {\cal L}_{10} = 2h_1 \zeta_{\alpha(3)} [ \pi^{\alpha(2)}
\rho^\alpha - \varphi \rho^{\alpha(3)} ]. 
\end{equation}
Collecting all variations together
\begin{eqnarray}
\delta {\cal L}_1 &=& \zeta_{\alpha(3)} [ - 2M_0g_1 
\pi^{\alpha(3)\beta} \rho_\beta + 2M^2g_1 \varphi \rho^{\alpha(3)}]
\nonumber \\
 && + \zeta_{\alpha(3)} [- f_2 \pi^{\alpha(3)\beta}
\rho_\beta + M_0f_2 \pi^{\alpha(2)} \rho^\alpha ] \nonumber \\
 && + 2h_1 \zeta_{\alpha(3)} [ \pi^{\alpha(2)}
\rho^\alpha - \varphi \rho^{\alpha(3)} ]
\end{eqnarray}
we obtain
$$
f_2 = - 2M_0g_1, \qquad h_1 = M_0{}^2g_1, \qquad
0 = 2M^2g_1 - 2h_1 = 2(M^2-M_0{}^2)g_1.
$$
From the last relation we see that we have two solutions 
$M_0 = \pm M$. Recall that in three dimensions the massive boson has
two physical degrees of freedom corresponding to two helicities $\pm
s$, while the massive fermion has only one physical degree of freedom
corresponding to helicity $+ s$ or $-s$. Moreover, the sign of the
helicity corresponds to the sign of the mass term in the free
Lagrangian. Thus the two solutions obtained correspond to two possible
helicities of the massive spinor.

So we have constructed the cubic vertex
\begin{equation}
{\cal L}_1 = g_1 \Psi_{\alpha(3)} [ (e^{\alpha(2)} \pi^{\alpha\beta}
- 2 e^{\alpha\beta} \pi^{\alpha(2)}) D \rho_\beta - 2M_0
E^{\alpha\beta} \pi^{\alpha(2)} \rho_\beta + M_0{}^2 E^{\alpha(2)}
\varphi \rho^\alpha ], 
\end{equation}
which is invariant under the following local hypertransformations: 
\begin{equation}
\delta \rho^\alpha = - 4g_1 \pi^{\alpha\beta(3)} \zeta_{\beta(3)} -
12M_0g_1 \pi_{\beta(2)} \zeta^{\alpha\beta(2)}, \qquad 
\delta \varphi = 4g_1 \rho^{\alpha(3)} \zeta_{\alpha(3)}. 
\end{equation}

\subsection{Vertex $(\frac{5}{2}-\frac{3}{2}-0)$}

Here we begin with the same terms with two derivatives as before
\begin{equation}
{\cal L}_{12} = g_1 \Psi_{\alpha(3)} [ e^{\alpha(2)} \pi^{\alpha\beta}
- 2 e^{\alpha\beta} \pi^{\alpha(2)} ] D \rho_\beta.
\end{equation}
The variations are the same, but one has take into account that the 
$\rho^\alpha$ equation is different. So, after introduction of the
hypertransformations
\begin{equation}
\delta_1 \rho^\alpha = - 2g_1 \pi^{\alpha\beta(3)} \zeta_{\beta(3)},
\qquad \delta_1 \varphi = 4g_1 \rho^{\alpha(3)} \zeta_{\alpha(3)}, 
\end{equation}
we obtain
\begin{equation}
\delta_0 {\cal L}_{12} + \delta_1 {\cal L}_0 = g_1 \zeta_{\alpha(3)} 
[ \pi^{\alpha(3)\beta} (12M_0 \phi_\beta - 6M_0 \rho_\beta)  + 2M^2
\varphi \rho^{\alpha(3)}],  \label{var2}
\end{equation}
where zero-form $\phi^\alpha$ comes from
$$
\Phi^\alpha = e_{\beta(2)} \phi^{\alpha\beta(2)} + e^\alpha{}_\beta
\phi^\beta.
$$
Now we introduce terms with one derivative
\begin{equation}
{\cal L}_{11} = \Psi_{\alpha(3)} [ f_1 (e^{\alpha(2)}
\pi^{\alpha\beta} - 2 e^{\alpha\beta} \pi^{\alpha(2)}) \Phi_\beta
+ f_2 E^{\alpha\beta} \pi^{\alpha(2)} \rho_\beta ],
\end{equation}
which produce variations
\begin{eqnarray}
\delta_0 {\cal L}_{11} &=& \zeta_{\alpha(3)} [ 12f_1 (D\pi)
\phi^{\alpha(3)} + 6f_1 \pi^{\alpha(3)\beta} \phi_\beta - f_2
\pi^{\alpha(3)\beta} \rho_\beta] \nonumber \\
 && + \zeta_{\alpha(3)} [ (f_1 e^{\alpha(2)} \pi^{\alpha\beta} - 2f_1
e^{\alpha\beta} \pi^{\alpha(2)}) D \Phi_\beta + 2f_3 \pi^{\alpha(2)}
D^\alpha{}_\beta \rho^\beta].
\end{eqnarray}
Then we add the following hypertransformations
\begin{eqnarray}
\delta_2 \Phi^\alpha &=& f_1 (2 e^{\alpha\beta} \pi^{\beta(2)}
- e^{\beta(2)} \pi^{\alpha\beta}) \zeta_{\beta(3)}, \nonumber \\
\delta_2 \rho^\alpha &=& 3f_2 \pi_{\beta(2)} \zeta^{\alpha\beta(2)},
\qquad \delta_2 \varphi = 12f_1 \phi^{\alpha(3)} \zeta_{\alpha(3)}
\end{eqnarray}
and obtain
\begin{eqnarray}
\delta_0 {\cal L}_{11} + \delta_2 {\cal L}_0 &=& \zeta_{\alpha(3)} [
6f_1 \pi^{\alpha(3)\beta}
\phi_\beta - f_2 \pi^{\alpha(3)\beta} \rho_\beta ] \nonumber \\
 && + \zeta_{\alpha(3)} [ - 4M_0f_1 \pi^\alpha{}_\beta
\phi^{\alpha(2)\beta} + (8M_0f_1-3M_0f_2) \pi^{\alpha(2)}
( 2 \phi^\alpha - \rho^\alpha) ] \nonumber \\
 && + 6M^2f_1 \zeta_{\alpha(3)} \varphi \phi^{\alpha(3)}. \label{var1}
\end{eqnarray}
At last we add terms without derivatives
\begin{equation}
{\cal L}_{10} = \Psi_{\alpha(3)} [ h_1 e^{\alpha(2)} \varphi
\Phi^\alpha + h_2 E^{\alpha(2)} \varphi \rho^\alpha ],
\end{equation}
which give
\begin{equation}
\delta_0 {\cal L}_{10} = \zeta_{\alpha(3)} [ 2h_1 E^\alpha{}_\beta
\pi^{\alpha\beta} \Phi^\alpha + 2h_2 \pi^{\alpha(2)} \rho^\alpha
+ h_1 e^{\alpha(2)} \varphi D \Phi^\alpha - 2h_2 \varphi
\rho^{\alpha(3)} ].
\end{equation}
Thus we have to add the final part of the hypertransformations
\begin{equation}
\delta_3 \Phi^\alpha = 3h_1 \varphi e_{\beta(2)}
\zeta^{\alpha\beta(2)}
\end{equation}
and we obtain
\begin{eqnarray}
\delta_0 {\cal L}_{10} + \delta_3 {\cal L}_0 &=& \zeta_{\alpha(3)} 
[ - 4h_1 (2 \pi^\alpha{}_\beta \phi^{\alpha(2)\beta} + \pi^{\alpha(2)}
\phi^\alpha) + 2h_2 \pi^{\alpha(2)} \rho^\alpha \nonumber \\
 && \qquad  - 2h_2 \varphi \rho^{\alpha(3)} + 12h_1M_0  \varphi
\phi^{\alpha(3)} ]. \label{var0}
\end{eqnarray}
All variations (\ref{var2}) + (\ref{var1}) + (\ref{var0}) cancel
provided
\begin{equation}
f_1 = - 2M_0g_1, \qquad f_2 = - 6M_0g_1, \qquad
h_1 = h_2 = M_0{}^2g_1, \qquad M^2 = M_0{}^2.
\end{equation}
Thus we again obtain two solutions $M_0 = \pm M$ with the cubic vertex
\begin{eqnarray}
{\cal L}_1 &=& g_1 \Psi_{\alpha(3)} [ e^{\alpha(2)} \pi^{\alpha\beta}
- 2 e^{\alpha\beta} \pi^{\alpha(2)} ] D \rho_\beta \nonumber \\
 && + M_0g_1 \Psi_{\alpha(3)} [ - 2 (e^{\alpha(2)}
\pi^{\alpha\beta} - 2 e^{\alpha\beta} \pi^{\alpha(2)}) \Phi_\beta
- 6 E^{\alpha\beta} \pi^{\alpha(2)} \rho_\beta ] \nonumber \\
 && + M_0{}^2g_1 \Psi_{\alpha(3)} [  e^{\alpha(2)} \varphi
\Phi^\alpha +  E^{\alpha(2)} \varphi \rho^\alpha ]
\end{eqnarray}
which is invariant under the following local hypertransformations:
\begin{eqnarray}
\delta \Phi^\alpha &=& - 2M_0g_1 (2 e^{\alpha\beta} \pi^{\beta(2)}
- e^{\beta(2)} \pi^{\alpha\beta}) \zeta_{\beta(3)} 
+ 3M_0{}^2g_1 \varphi e_{\beta(2)} \zeta^{\alpha\beta(2)},\nonumber \\
\delta \rho^\alpha &=& - 2g_1 \pi^{\alpha\beta(3)} \zeta_{\beta(3)}
- 18M_0g_1 \pi_{\beta(2)} \zeta^{\alpha\beta(2)}, \\
\delta \varphi &=& 4g_1 \rho^{\alpha(3)} \zeta_{\alpha(3)}
-  24M_0g_1 \phi^{\alpha(3)} \zeta_{\alpha(3)}. \nonumber
\end{eqnarray}
{\bf Supersymmetry} Note that in our formalism the massive 
spin-$\frac{3}{2}$ field is also the gauge field with its own local
gauge transformations
$$
\delta \Phi^\alpha = D \zeta^\alpha + M_0 e^\alpha{}_\beta
\zeta^\beta, \qquad \delta \rho^\alpha = 2M_0 \zeta^\alpha.
$$
By direct calculations one can check the the vertex obtained is also
invariant under these supertransformations with the following
correction
\begin{equation}
\delta \Psi^{\alpha(3)} = h_1 \varphi e^{\alpha(2)} \zeta^\alpha . 
\end{equation}

\subsection{Vertex $(\frac{5}{2}-1-\frac{1}{2})$}

In this case, having already some experience, we begin with the
correct structure of the two derivative terms:
\begin{equation}
{\cal L}_{12} = \Psi_{\alpha(3)} [ g_1 (e^{\alpha(2)} B^{\alpha\beta}
- 2 e^{\alpha\beta} B^{\alpha(2)}) + g_2 (e^{\alpha(2)}
\pi^{\alpha\beta} - 2 e^{\alpha\beta} \pi^{\alpha(2)}) ] D \rho_\beta.
\label{lag31}
\end{equation}
They produce variations
\begin{eqnarray}
\delta_0 {\cal L}_{12} &=& \zeta_{\alpha(3)} 
[ (4g_1 B^{\alpha(3)\beta} + 4g_2 \pi^{\alpha(3)\beta}) 
D_\beta{}^\gamma \rho_\gamma   2g_1 (DB)^{\alpha\alpha}
D^{\alpha\beta} \rho_\beta - 4g_1 (DB)^{\alpha\beta} D^{\alpha(2)}
\rho_\beta ] \nonumber \\
 && + 4g_2 \zeta_{\alpha(3)} (D\pi) \rho^{\alpha(3)} 
- 2Mg_2 \zeta_{\alpha(3)} [ E^{\alpha(2)} B^{\alpha\beta}
- \frac{1}{2} E^{\alpha\beta} B^{\alpha(2)}] D \rho_\beta.
\end{eqnarray}
Now we introduce hypertransformations
\begin{eqnarray}
\delta_1 A^{\alpha(2)} &=& 3g_1 (\zeta^{\alpha(2)\beta} 
D_\beta{}^\gamma \rho_\gamma + \zeta^{\alpha\beta(2)} D_{\beta(2)}
\rho^\alpha ),   \nonumber \\
\delta_1 \rho^\alpha &=& - 4 (g_1 B^{\alpha\beta(3)} + g_2
\pi^{\alpha\beta(3)}) \zeta_{\beta(3)}, \\
\delta_1 \varphi &=& 4g_2 \rho^{\alpha(3)} \zeta_{\alpha(3)} \nonumber
\end{eqnarray}
and obtain
\begin{eqnarray}
\delta_0 {\cal L}_{12} + \delta_1 {\cal L}_0 &=& \zeta_{\alpha(3)} 
[ - 2M_0g_1 B^{\alpha(3)\beta} - 2M_0g_2 \pi^{\alpha(3)\beta}]
\rho_\beta \nonumber \\
 && + \zeta_{\alpha(3)} [ - 2Mg_2 E^{\alpha(2)} B^{\alpha\beta} + Mg_2
E^{\alpha\beta} B^{\alpha(2)} \nonumber \\
 && \qquad - 2Mg_1 E^{\alpha(2)} \pi^{\alpha\beta}
+ Mg_1 E^{\alpha\beta} \pi^{\alpha(2)}] D \rho_\beta.
\end{eqnarray}
We proceed with the terms with one derivative
\begin{equation}
{\cal L}_{11} = \Psi_{\alpha(3)} [ f_1 E^{\alpha(2)} B^{\alpha\beta}
+ f_2 E^{\alpha\beta} B^{\alpha(2)} + f_3 E^{\alpha(2)}
\pi^{\alpha\beta} + f_4 E^{\alpha\beta} \pi^{\alpha(2)}] \rho_\beta
\label{lag32}
\end{equation}
which give variations
\begin{eqnarray}
\delta_0 {\cal L}_{11} &=& \zeta_{\alpha(3)} [ - (f_1+f_2)
B^{\alpha(3)\beta} - (f_3+f_4) \pi^{\alpha(3)\beta}] \rho_\beta
\nonumber \\
 && + \zeta_{\alpha(3)} [ - \frac{f_1-f_2}{2} D^\alpha{}_\beta 
B^{\alpha\beta} \rho^\alpha - \frac{f_3-f_4}{2}M B^{\alpha(2)}
\rho^\alpha ] \nonumber \\
 &&  - \zeta_{\alpha(3)} [ f_1 E^{\alpha(2)} B^{\alpha\beta}
+ f_2 E^{\alpha\beta} B^{\alpha(2)} + f_3 E^{\alpha(2)}
\pi^{\alpha\beta} + f_4 E^{\alpha\beta} \pi^{\alpha(2)}] D \rho_\beta.
\end{eqnarray}
So we add
\begin{equation}
\delta_2 A^{\alpha(2)} = - \frac{3}{4}(f_1-f_2) \zeta^{\alpha(2)\beta}
\rho_\beta
\end{equation}
and obtain
\begin{eqnarray}
\delta_0 {\cal L}_{11} + \delta_2 {\cal L}_0 &=& \zeta_{\alpha(3)} 
[ - (f_1+f_2) B^{\alpha(3)\beta} - (f_3+f_4) \pi^{\alpha(3)\beta} ]
\rho_\beta \nonumber \\
 && - \zeta_{\alpha(3)} [ f_1 E^{\alpha(2)} B^{\alpha\beta}
+ f_2 E^{\alpha\beta} B^{\alpha(2)} + f_3 E^{\alpha(2)}
\pi^{\alpha\beta} + f_4 E^{\alpha\beta} \pi^{\alpha(2)}] D \rho_\beta
\nonumber \\
 && + \zeta_{\alpha(3)} [ - \frac{M}{2}(f_3-f_4) B^{\alpha(2)}
\rho^\alpha - \frac{M}{2}(f_1-f_2) \pi^{\alpha(2)} \rho^\alpha ]. 
\end{eqnarray}
We have to put
\begin{equation}
f_1 = - 2Mg_2, \quad f_2 = 2Mg_2 - 2M_0g_1, \quad
f_3 = - 2Mg_1, \quad f_4 = 2Mg_1 - 2M_0g_2,
\end{equation}
then total variations reduce to
\begin{eqnarray}
\Delta &=& \zeta_{\alpha(3)} [  2(Mg_2-2M_0g_1) B^{\alpha(2)} + 
2(Mg_1-2M_0g_2) \pi^{\alpha(2)} ] D^\alpha{}_\beta \rho^\beta
\nonumber \\
 && + \zeta_{\alpha(3)} [ M( 2Mg_1-M_0g_2) B^{\alpha(2)}
\rho^\alpha + M(2Mg_2-M_0g_1) \pi^{\alpha(2)} \rho^\alpha ].
\end{eqnarray}
Note that the gauge invariance of the vector field forbids any terms
without derivatives, so the last possibility to cancel the variations
is to introduce
\begin{equation}
\delta_3 \rho^\alpha = 6[ (f_2 - Mg_2) B_{\beta(2)} 
+ (f_4 - Mg_1)\pi_{\beta(2)} ] \zeta^{\alpha\beta(2)}.
\label{var32}
\end{equation}
As in the previous cases we need $M^2 = M_0{}^2$. Thus we obtain the
vertex  
\begin{eqnarray}
{\cal L}_1 &=& \Psi_{\alpha(3)} [ g_1 e^{\alpha(2)} B^{\alpha\beta} -
2g_1 e^{\alpha\beta} B^{\alpha(2)} + g_2 e^{\alpha(2)}
\pi^{\alpha\beta} - 2g_2 e^{\alpha\beta} \pi^{\alpha(2)}] D \rho_\beta
\nonumber \\
 && + \Psi_{\alpha(3)} [ f_1 E^{\alpha(2)} B^{\alpha\beta}
+ f_2 E^{\alpha\beta} B^{\alpha(2)} + f_3 E^{\alpha(2)}
\pi^{\alpha\beta} + f_4 E^{\alpha\beta} \pi^{\alpha(2)}] \rho_\beta
\end{eqnarray}
which is invariant under the following local hypertransformations
\begin{eqnarray}
\delta A^{\alpha(2)} &=& 3g_1 (\zeta^{\alpha(2)\beta} D_\beta{}^\gamma
\rho_\gamma + \zeta^{\alpha\beta(2)} D_{\beta(2)} \rho^\alpha )
- \frac{3}{4}(f_1-f_2) \zeta^{\alpha(2)\beta} \rho_\beta, \nonumber \\
\delta \rho^\alpha &=& - 4(g_1 B^{\alpha\beta(3)} + g_2
\pi^{\alpha\beta(3)} ) \zeta_{\beta(3)} \\
 && + 6[ (f_2 - Mg_2) B_{\beta(2)} + (f_4 - Mg_1)\pi_{\beta(2)}
] \zeta^{\alpha\beta(2)}, \nonumber \\
\delta \varphi &=& 4g_2 \rho^{\alpha(3)} \zeta_{\alpha(3)}. \nonumber
\end{eqnarray}

\subsection{Vertex $(\frac{5}{2}-\frac{3}{2}-1)$}

In this case the gauge invariance of the vector field also forbids any
terms without derivatives, so the vertex contain three parts:
\begin{equation}
{\cal L}_1 = {\cal L}_{12} + {\cal L}_{11a} + {\cal L}_{11b},
\end{equation}
\begin{eqnarray}
{\cal L}_{12} &=& \Psi_{\alpha(3)} [ g_1 (e^{\alpha(2)}
B^{\alpha\beta} - 2 e^{\alpha\beta} B^{\alpha(2)}) + g_2
(e^{\alpha(2)} \pi^{\alpha\beta} - 2 e^{\alpha\beta} \pi^{\alpha(2)})
] D \rho_\beta,  \\
{\cal L}_{11a} &=& \Psi_{\alpha(3)} [ f_1 E^{\alpha(2)}
B^{\alpha\beta} + f_2 E^{\alpha\beta} B^{\alpha(2)} + f_3
E^{\alpha(2)} \pi^{\alpha\beta} + f_4 E^{\alpha\beta} \pi^{\alpha(2)}]
\rho_\beta,  \\
{\cal L}_{11b} &=& \Psi_{\alpha(3)} [ f_5 (e^{\alpha(2)}
B^{\alpha\beta} - 2 e^{\alpha\beta} B^{\alpha(2)}) + f_6
(e^{\alpha(2)} \pi^{\alpha\beta} - 2 e^{\alpha\beta} \pi^{\alpha(2)})
] \Phi_\beta,
\end{eqnarray}
where ${\cal L}_{12}$ and ${\cal L}_{11a}$ are the same as before,
while ${\cal L}_{11b}$ is new. Variations of the last one
\begin{eqnarray}
\delta_0 {\cal L}_{11b} &=& \zeta_{\alpha(3)} [2f_5 B^{\alpha(3)\beta}
+ 2f_6 \pi^{\alpha(3)\beta}] E_\beta{}^\gamma \Phi_\gamma \nonumber \\
 && + \zeta_{\alpha(3)} [ - 2f_5 E^{\alpha(2)} (DB)^{\alpha\beta}
+ f_5 E^{\alpha\beta} (DB)^{\alpha\alpha})] \Phi_\beta \nonumber \\
 && + \zeta_{\alpha(2)} [ - 2Mf_6 E^{\alpha(2)} B^{\alpha\beta}
+ Mf_6 E^{\alpha\beta} B^{\alpha(2)} ] \Phi_\beta \nonumber \\
 && + \zeta_{\alpha(3)} [ f_5 (e^{\alpha(2)} B^{\alpha\beta}
- 2 e^{\alpha\beta} B^{\alpha(2)}) + f_6 (e^{\alpha(2)}
\pi^{\alpha\beta} - 2 e^{\alpha\beta} \pi^{\alpha(2)}) ] D\Phi_\beta
\end{eqnarray}
require the following hypertransformations:
\begin{eqnarray}
\delta \Phi^\alpha &=& [ f_5 (e^{\beta(2)} B^{\alpha\beta} - 2
e^{\alpha\beta} B^{\beta(2)}) + f_6 (e^{\beta(2)} \pi^{\alpha\beta}
- 2 e^{\alpha\beta} \pi^{\beta(2)}) ] \zeta_{\beta(3)}, \nonumber \\
\delta A^{\alpha(2)} &=& 3f_5 [ \zeta^{\alpha\beta(2)} 
\phi^\alpha{}_{\beta(2)} - \frac{5}{2} \zeta^{\alpha(2)\beta}
\phi_\beta ], \\
\delta \varphi &=& 4f_6 \phi^{\alpha(3)} \zeta_{\alpha(3)}. \nonumber
\end{eqnarray}
All other calculations look similar to the ones above, so we present
only final results here. Once again we obtain two solutions, this time
$M_0 = \pm M$, $g_1 = \pm g_2$. All other coefficients are
\begin{equation}
f_1 = f_5 = - 2Mg_2, \quad f_2 = - 4Mg_2, \quad
f_3 = f_6 = - 2Mg_1, \quad f_4 = - 4Mg_1.
\end{equation}
A complete set of the hypertransformations has the form:
\begin{eqnarray}
\delta \Phi^\alpha &=& - 2M [ g_2 (e^{\beta(2)} B^{\alpha\beta} - 2
e^{\alpha\beta} B^{\beta(2)}) + g_1 (e^{\beta(2)} \pi^{\alpha\beta}
- 2 e^{\alpha\beta} \pi^{\beta(2)}) ] \zeta_{\beta(3)}, \nonumber \\
\delta A^{\alpha(2)} &=& 3g_1 (\zeta^{\alpha(2)\beta} D_\beta{}^\gamma
\rho_\gamma + \zeta^{\alpha\beta(2)} D_{\beta(2)} \rho^\alpha )
- \frac{3Mg_2}{2} \zeta^{\alpha(2)\beta} \rho_\beta \nonumber  \\
 && - 3Mg_2 [ 2 \zeta^{\alpha\beta(2)} \phi^\alpha{}_{\beta(2)} - 5
\zeta^{\alpha(2)\beta} \phi_\beta ],   \\
 \delta \rho^\alpha &=& - 2 [g_1 B^{\alpha\beta(3)} + g_2
\pi^{\alpha\beta(3)}] \zeta_{\beta(3)} + 15M [g_2 B_{\beta(2)} +
g_1 \pi_{\beta(2)}] \zeta^{\alpha\beta(2)}, \nonumber  \\
\delta \varphi &=& 4g_2 \rho^{\alpha(3)}, \zeta_{\alpha(3)}
- 8Mg_1 \phi^{\alpha(3)} \zeta_{\alpha(2)}. \nonumber
\end{eqnarray}
We have also explicitly checked that the vertex is  invariant under
the supertransformations of the massive spin-$\frac{3}{2}$ field.

\subsection{Vertex $(\frac{5}{2}-2-\frac{1}{2})$}

A complete ansatz for the cubic vertex (including the terms without
derivatives which are now allowed) has the form
\begin{equation}
{\cal L}_1 = {\cal L}_{12} + {\cal L}_{11} + {\cal L}_{10},
\end{equation}
\begin{eqnarray}
{\cal L}_{12} &=& \Psi_{\alpha(3)} [ g_1 (e^{\alpha(2)}
B^{\alpha\beta} - 2 e^{\alpha\beta} B^{\alpha(2)}) + g_2
(e^{\alpha(2)} \pi^{\alpha\beta} - 2 e^{\alpha\beta} \pi^{\alpha(2)})]
D \rho_\beta, \\
{\cal L}_{11} &=&  f_0 \Psi_{\alpha(3)} ( e^{\alpha(2)}
\Omega^{\alpha\beta} - 2 e^{\alpha\beta} \Omega^{\alpha(2)})
\rho_\beta \nonumber \\
 && + \Psi_{\alpha(3)} [ f_1 E^{\alpha(2)} B^{\alpha\beta} + f_2
E^{\alpha\beta} B^{\alpha(2)} + f_3 E^{\alpha(2)} \pi^{\alpha\beta}
+ f_4 E^{\alpha\beta} \pi^{\alpha(2)} ] \rho_\beta, \\
{\cal L}_{10} &=& \Psi_{\alpha(3)} [ h_0 (e^{\alpha(2)}
H^{\alpha\beta} - 2 e^{\alpha\beta} H^{\alpha(2)}) \rho_\beta + h_1
E^{\alpha(2)} \varphi \rho^\alpha ]. 
\end{eqnarray}
As one can see the only new terms are the ones with the spin-2
components $\Omega^{\alpha(2)}$ and $H^{\alpha(2)}$. After rather long
calculations we omit here we again have found that there are two
solutions $M_0 = \pm M$, $g_1 = \pm g_2$, while hypersymmetry requires
$$
f_1 = - 4Mg_2, \qquad f_2 = 2Mg_2, \qquad
f_3 = - 4Mg_1, \qquad f_4 = 2Mg_1, 
$$
$$
f_0 = 2Mg_1, \qquad h_0 = M^2g_1, \qquad h_1 = 3h_0. 
$$
A complete set of the hypertransformations for the one-forms looks
like
\begin{eqnarray}
\delta \Omega^{\alpha(2)} &=& \frac{3h_0}{2} ( 2 
\zeta^{\alpha(2)\beta} e_{\beta(2)} \rho^\beta -
\zeta^{\alpha\beta(2)} e_{\beta(2)} \rho^\alpha ), \nonumber \\
\delta H^{\alpha(2)} &=& \frac{3f_0}{2} ( 2 
\zeta^{\alpha(2)\beta} e_{\beta(2)} \rho^\beta -
\zeta^{\alpha\beta(2)} e_{\beta(2)} \rho^\alpha ), 
\end{eqnarray}
while for the zero-forms we obtain
\begin{eqnarray}
\delta A^{\alpha(2)} &=& 3g_1 (\zeta^{\alpha(2)\beta} D_\beta{}^\gamma
\rho_\gamma + \zeta^{\alpha\beta(2)} D_{\beta(2)} \rho^\alpha )
- \frac{9Mg_2}{4} \zeta^{\alpha(2)\beta} \rho_\beta, \nonumber \\
\delta \rho^\alpha &=& - 4 (g_1 B^{\alpha\beta(3)} + g_2
\pi^{\alpha\beta(3)}) \zeta_{\beta(3)} + 24Mg_1 
(2 \omega^{\alpha\beta(3)} + M h^{\alpha\beta(3)}) \zeta_{\beta(3)} 
\nonumber \\
 && + 6 [Mg_2 B_{\beta(2)} + Mg_1 \pi_{\beta(2)}]
\zeta^{\alpha(2)\beta}, \\
\delta \varphi &=& 2g_2 \rho^{\alpha(3)} \zeta_{\alpha(3)}. \nonumber 
\end{eqnarray}
We also have explicitly checked that the vertex is invariant under
the massive spin-2 gauge transformations where massless 
spin-$\frac{5}{2}$ transforms as follows:
\begin{equation}
\delta \Psi^{\alpha(3)} = 2M [ g_1(e^{\alpha(2)} \eta^{\alpha\beta}
- 2 e^{\alpha\beta} \eta^{\alpha(2)}) + g_2 (e^{\alpha(2)}
\xi^{\alpha\beta} - 2 e^{\alpha\beta} \xi^{\alpha(2)})] \rho_\beta. 
\end{equation}

\subsection{Vertex $(\frac{5}{2},2,\frac{3}{2})$}

Here a complete ansatz for the cubic vertex has the form
\begin{eqnarray}
{\cal L}_1 &=& \Psi_{\alpha(3)} [ g_1 e^{\alpha(2)} B^{\alpha\beta} -
2g_1 e^{\alpha\beta} B^{\alpha(2)} + g_2 e^{\alpha(2)}
\pi^{\alpha\beta} - 2g_2 e^{\alpha\beta} \pi^{\alpha(2)}] D \rho_\beta
\nonumber \\
 && + \Psi_{\alpha(3)} [ \tilde{f}_0 \Omega^{\alpha(2)} \Phi^\alpha  
+ f_0 ( e^{\alpha(2)} \Omega^{\alpha\beta} - 2 e^{\alpha\beta}
\Omega^{\alpha(2)}) \rho_\beta ] \nonumber \\
 && + \Psi_{\alpha(3)} [ f_5 (e^{\alpha(2)} B^{\alpha\beta}
- 2 e^{\alpha\beta} B^{\alpha(2)} + f_6 (e^{\alpha(2)}
\pi^{\alpha\beta} - 2 e^{\alpha\beta} \pi^{\alpha(2)})] \Phi_\beta
\nonumber \\
 && + \Psi_{\alpha(3)} [ f_1 E^{\alpha(2)}
B^{\alpha\beta} + f_2 E^{\alpha\beta} B^{\alpha(2)} + f_3
E^{\alpha(2)} \pi^{\alpha\beta} + f_4 E^{\alpha\beta} \pi^{\alpha(2)}]
\rho_\beta \nonumber \\
 && + \Psi_{\alpha(3)} [ \tilde{h}_0 H^{\alpha(2)} \Phi^\alpha 
+ h_0 (e^{\alpha(2)} H^{\alpha\beta} - 2 e^{\alpha\beta}
H^{\alpha(2)}) \rho_\beta + \tilde{h}_1 e^{\alpha(2)} \varphi
\Phi^\alpha + h_1 E^{\alpha(2)} \varphi \rho^\alpha]. 
\end{eqnarray}
Most of these terms we are already familiar with, while the only new
ones are the terms with the coefficients $\tilde{f}_0$ and 
$\tilde{h}_0$. In this case the hypersymmetry requires
$$
M_0 = \pm M, \quad g_1 = \pm g_2, \qquad
\tilde{f}_0 = 4Mg_1, \qquad f_0 = 2Mg_1, 
$$
$$
f_1 = - 4Mg_2, \qquad f_2 = f_5 = - 2Mg_2, \quad
f_3 = - 4Mg_1, \qquad f_4 = f_6 = - 2Mg_1,
$$
$$
\tilde{h}_0 = 2M^2g_2, \qquad h_0 = M^2g_2, \qquad
\tilde{h}_1 = M^2g_2, \qquad h_1 = 3M^2g_2.
$$
In this, a complete set of the hypertransformations for the one-forms
looks like
\begin{eqnarray}
\delta \Omega^{\alpha(2)} &=& - 3\tilde{h}_0 \zeta^{\alpha(2)\beta}
\Phi_\beta + \frac{3h_0}{2} ( 2 \zeta^{\alpha(2)\beta} e_{\beta(2)}
\rho^\beta - \zeta^{\alpha\beta(2)} e_{\beta(2)} \rho^\alpha ),
\nonumber \\
\delta H^{\alpha(2)} &=& - 3\tilde{f}_0 \zeta^{\alpha(2)\beta}
\Phi_\beta + \frac{3f_0}{2} ( 2 \zeta^{\alpha(2)\beta} e_{\beta(2)}
\rho^\beta - \zeta^{\alpha\beta(2)} e_{\beta(2)} \rho^\alpha ),  \\
\delta \Phi^\alpha &=& 3\tilde{f}_0 \zeta^{\alpha\beta(2)}
\Omega_{\beta(2)} + 3\tilde{h}_0 \zeta^{\alpha\beta(2)} H_{\beta(2)} 
+ 3\tilde{h}_1 \varphi e_{\beta(2)} \zeta^{\alpha\beta(2)}
\nonumber \\
 && + [f_5 (e^{\beta(2} B^{\alpha\beta} - 2 e^{\alpha\beta}
B^{\beta(2)}) + f_6 (e^{\beta(2)} \pi^{\alpha\beta} - 2
e^{\alpha\beta} \pi^{\beta(2)}) ] \zeta_{\beta(3)}, \nonumber
\end{eqnarray}
while for the zero-forms we obtain
\begin{eqnarray}
\delta A^{\alpha(2)} &=& 3g_1 (\zeta^{\alpha(2)\beta} D_\beta{}^\gamma
\rho_\gamma + \zeta^{\alpha\beta(2)} D_{\beta(2)} \rho^\alpha) 
- \frac{3}{8}(f_1-f_2) \zeta^{\alpha(2)\beta} \rho_\beta  
\nonumber \\
&& + 3f_5 [ \zeta^{\alpha\beta(2)} \phi^\alpha{}_{\beta(2)} -
\frac{5}{2} \zeta^{\alpha(2)\beta} \phi_\beta], \nonumber \\
\delta \rho^\alpha &=& - 2(g_1 B^{\alpha\beta(3)} + g_2
\pi^{\alpha\beta(3)}) \zeta_{\beta(3)} + 24Mg_1 (2
\omega^{\alpha\beta(3)} + M h^{\alpha\beta(3)}) \zeta_{\beta(3)} \\
 && +  4M[g_2 B_{\beta(2)} + Mg_1 \pi_{\beta(2)}]
\zeta^{\alpha\beta(2)}, \nonumber \\ 
\delta \varphi &=& 2g_2 \rho^{\alpha(3)} \zeta_{\alpha(3)}
+ 2f_6 \phi^{\alpha(3)} \zeta_{\alpha(3)}.  \nonumber 
\end{eqnarray}
The vertex is also invariant both under the massive spin-2 gauge
transformations with 
\begin{eqnarray}
\delta \Psi^{\alpha(3)} &=& \tilde{f}_0 \eta^{\alpha(2)} \Phi^\alpha
+ \tilde{h}_0 \xi^{\alpha(2)} \Phi^\alpha ,\nonumber \\
\delta \Phi^\alpha &=& 3\tilde{f}_0 \eta_{\beta(2)}
\Psi^{\alpha\beta(2)} + 3\tilde{h}_0 \xi_{\beta(2)}
\Psi^{\alpha\beta(2)} 
\end{eqnarray}
as well as under the local supersymmetry of the massive 
spin-$\frac{3}{2}$ field
\begin{eqnarray}
\delta \Psi^{\alpha(3)} &=& \tilde{f}_0 \Omega^{\alpha(2)}
\zeta^\alpha + \tilde{h}_0 H^{\alpha(2)} \zeta^\alpha, \nonumber \\
\delta \Omega^{\alpha(2)} &=& 3\tilde{h}_0 \Psi^{\alpha(2)\beta}
\zeta_\beta, \qquad \delta H^{\alpha(2)} = 3\tilde{f}_0 
\Psi^{\alpha(2)\beta} \zeta_\beta. 
\end{eqnarray}
And this serves as a non-trivial check for the consistency of the
whole construction.

\section*{Conclusion}

In this work we presented a number of explicit examples for the cubic
vertices describing an interaction of massless spin-$\frac{5}{2}$
field with massive boson and fermion including all
hypertransformations necessary for the vertices to be gauge invariant.
Here we restrict ourselves with the massive bosons with spins
$s=2,1,0$ and massive fermions with spins $s=\frac{3}{2},\frac{1}{2}$.
Our general analysis \cite{Zin21} predicted that the vertex must exist
for any boson and fermion with the spin difference $\frac{3}{2}$ or 
$\frac{1}{2}$. And indeed it appeared that the vertex exists for all
six possible pairs $(2,1,0) \otimes (\frac{3}{2},\frac{1}{2})$. As in
the case of massive supermultiplets \cite{BSZ17a,Zin23}, our
construction is based on the gauge invariant description for the
massive fields with spins $s \ge 1$. Moreover, we have explicitly
checked that all the vertices are invariant also under the gauge
symmetries of these massive fields.

An open question is what is the hypermultiplet. Indeed, we have seen
that the whole hypergravity contains also a spin-4 field. And indeed,
if one calculates the anti-commutator of the hypertransformations
given above, one finds that they generate higher derivative
transformations corresponding to the spin-4 field. Thus to determine
what is a hypermultiplet, i.e. the minimal set of fields necessary to
realize the whole superalgebra, we have to consider spin-4
interactions as well. We leave this task for the future work.

\end{document}